# ELECTRON STATES IN A STAIRCASE INFINITELY DEEP WELL IN A MAGNETIC FIELD

M.M. AGHASYAN and A.A. KIRAKOSYAN

The electron states in a size-quantaized coated semiconductor wire in a uniform magnetic field applied parallel to the wire axis is considered. The wave functions are found and the equation for determination of energy eigenvalues depending both on magnetic field and parameters of the wire and the coating is obtained. The binding energy of a hydrogen-like impurity located on the wire axis is computed by the use of the variational method, and it is shown that this energy increases with the increase of the magnetic field.

## 1. Introduction

Magnetic field effect on parameters of low-dimensional electronic systems is revealing in various physical phenomena, having both the fundamental meaning for physical theory [1], and various applications. The combination of the possibilities of modern experimental technique in obtaining magnetic fields up to 1000 T [2] and technologies in the making of low-dimensional systems with different geometric forms [3,4] opens broad perspective for controlling properties of such systems and creating on those basis new functional nanoelectronic setups.

The study of the influence of the magnetic field effect on electron states in a size-quantized semiconductor wire within the framework of the infinitely deep potential well model was studied in Refs. [5,6]. The same problem for the finite potential well model is shown in Ref. [7] where the binding energy of a charged impurity has been also calculated.

In this paper the energy spectrum and wave functions determination problem for electron within the framework of the staircase infinitely deep well model (SIW) suggested in [8], is considered in magnetic field directed along the axis of the wire. The binding energy of a hydrogen-like impurity located at the wire axis depending on the magnetic field value and characterizing parameters of the considered model is calculated.

## 2. Wave functions and energy spectrum

Let us consider the electron states in the wire of radius $R_1$, coated by the layer of thikness $R_2 - R_1$, in the magnetic field applied parallel to the wire axis (axis $z$). In the effective mass approximation method the Hamiltonian of this problem is of the form

$$\hat{H} = \frac{1}{2m_i}\left[\hat{\mathbf{p}} - \frac{e}{c}\mathbf{A}(\mathbf{r})\right]^2 + V(\mathbf{r}), \qquad (1)$$

where $m_i$ is the electron effective mass in the wire ($i=1$) and the layer ($i=2$), $\hat{\mathbf{p}}$ is the momentum operator, $\mathbf{A}(\mathbf{r})$ is the magnetic field vector potential. The potential energy $V(\mathbf{r})$ within the framework of SIW is of the form

$$V(\mathbf{r}) = \begin{cases} 0, & r < R_1, \\ V_0, & R_1 \leq r \leq R_2, \\ \infty, & r > R_2, \end{cases} \qquad (2)$$

where $V_0$ is the value of the potenial energy jump at the boundary of the wire and the coating layer. In the case of uniform magnetic field the vector potential $\mathbf{A}(\mathbf{r}) = [\mathbf{H}, \mathbf{r}]/2$, where $\mathbf{H} \equiv \mathbf{H}(0,0,H)$, and therefore in the cylindrical coordinate system only the component $\mathbf{A}_j = Hr/2$ differs from zero.

The solution of the Schroedinger equation with the Hamiltonian (1), where $V(\mathbf{r})$ is given by Eq. (2), can be presented in the form

$$\Psi_{nlk}(r, \boldsymbol{j}, z) = \frac{C_1}{a_c\sqrt{2pL}} \exp[i(l\boldsymbol{j} + kz)] x^{|l|/2} \exp\left(-\frac{x}{2}\right) \times$$

$$\times \begin{cases} F(-\boldsymbol{a}_{n,|l|}, |l|+1; x), & x < x_1, \\ C_2 F(-\boldsymbol{b}_{n,|l|}, |l|+1; x) + C_3 U(-\boldsymbol{b}_{n,|l|}, |l|+1; x), & x_1 \leq x \leq x_2, \\ 0, & x > x_2, \end{cases} \qquad (3)$$

where $L$ is the wire length, $k$ is the wave-number, $n = 0,1,2,...$, $l = 0, \pm 1, \pm 2,...$, are the quantum numbers, $F(a,b;x)$ and $U(a,b;x)$ are the confluente hypergeometric functions [9], $a_c = \sqrt{\hbar c/eH}$ is the magnetic length, $x = r^2/2a_c^2$. The normalization constants $C_j$ ($j=1,2,3$) entering in the solution (3) are given by the expressions

$$C_1 = \left\{ \int_0^{x_1} dx\, e^{-x} x^{|l|} F^2\left(-a_{n,|l|}, |l|+1; x\right) + \int_{x_1}^{x_2} dx\, e^{-x} x^{|l|} \left[C_2 F\left(-b_{n,|l|}, |l|+1; x\right) + C_3 U\left(-b_{n,|l|}, |l|+1; x\right)\right]^2 \right\}^{-\frac{1}{2}}$$
, (4)

$$C_2 = \frac{F(-a_{n,|l|}, |l|+1; x_1) U(-b_{n,|l|}, |l|+1; x_2)}{F(-b_{n,|l|}, |l|+1; x_1) U(-b_{n,|l|}, |l|+1; x_2) - F(-b_{n,|l|}, |l|+1; x_2) U(-b_{n,|l|}, |l|+1; x_1)}, \quad (5)$$

$$C_3 = -C_2 \frac{F(-b_{n,|l|}, |l|+1; x_2)}{U(-b_{n,|l|}, |l|+1; x_2)}, \quad (6)$$

where

$$x_1 = \frac{R_1^2}{2a_c^2}, \quad x_2 = \frac{R_2^2}{2a_c^2}. \quad (7)$$

The energy eigenvalues are given by the expression

$$E_{nlk} = \frac{\hbar^2 k^2}{2m_1} + \hbar w_c \left(a_{n,|l|} + \frac{|l|+l+1}{2}\right), \quad (8)$$

where $w_c = eH/m_1 c$ is the cyclotron frequency. The quantum numbers $b_{n,|l|}$ are expressed in $a_{n,|l|}$ from the condition

$$\hbar w_c \left(a_{n,|l|} + \frac{|l|+l+1}{2}\right) = V_0 + \hbar w_c \frac{m_1}{m_2}\left(b_{n,|l|} + \frac{|l|+l+1}{2}\right). \quad (9)$$

The quantum numbers $a_{n,|l|}$ are determined from the continuity condition of the logarithmic derivative of the wave functions at $r = R_1$ and are the roots of the equation

$$\frac{1}{m_1} \frac{d}{dx_1} \ln\left[e^{-\frac{x_1}{2}} x_1^{\frac{|l|}{2}} F\left(-a_{n,|l|}, |l|+1; x_1\right)\right] =$$

$$= \frac{1}{m_2} \frac{d}{dx_1} \ln\left\{e^{-\frac{x_1}{2}} x_1^{\frac{|l|}{2}} \left[C_2 F\left(-b_{n,|l|}, |l|+1; x_1\right) + C_3 U\left(-b_{n,|l|}, |l|+1; x_1\right)\right]^2\right\}. \quad (10)$$

**3. Calculation of binding energy**

We use the variational method to determine the binding energy of the hydrogen-like center located on the wire axis. The potential energy of electron in the impurity center field is given by the expression

$$U(r,z) = \frac{-e^2}{c\sqrt{r^2 + z^2}},  \quad (11)$$

where it is assumed that the wire and the coating have the same dielectric constants: $c_1 = c_2 \equiv c$.

Following Ref. [10] we shall choose the ground state trial wave function in the form

$$\Psi_0 = N\exp(ikz)\exp\left(-\frac{x}{2}\right)\exp\left(-l\sqrt{r^2+z^2}\right)\begin{cases} F(-a,1;x), & x < x_1, \\ C_2 F(-b,1;x) + C_3 U(-b,1;x), & x_1 \leq x \leq x_2, \\ 0, & x > x_2, \end{cases} \quad (12)$$

where $l$ is the variational parameter, $N$ is the normalization constant, $a \equiv a_{1,0}$, $b \equiv b_{1,0}$.

The binding energy ($E_b$) of impurity we determined as the difference of the ground state energy of the system without impurity ($E_{100}$) and the energy of the ground state with impurity ($E_0$):

$$E_b = E_{100} - E_0. \quad (13)$$

With the simple calculations for the hydrogen-like impurity binding energy we get the expression

$$\frac{E_b}{E_R} = -g^2 + 2\frac{f_0 + g_0}{f_1 + g_1} - g\frac{1 - m_1/m_2}{f_1 + g_1}\left[t_1^2 e^{-nt_1^2} F^2(-a,1;nt_1^2) K_0(2gt_1) - g\, g_1\right], \quad (14)$$

where the following notations are introduced: $g = l a_B$, $a_B$ and $E_R$ are the effective Bohr radius and Rydberg energy in the wire, respectively,

$$f_j(t_1, a, g, n) = \int_0^{t_1} e^{-nt^2} F^2(-a,1;nt^2) K_j(2gt) t^{j+1} dt, \quad (15)$$

$$g_j(t_1, t_2, b, g, n) = \int_{t_1}^{t_2} e^{-nt^2} \left[C_2 F(-b,1;nt^2) + C_3 U(-b,1;nt^2)\right]^2 K_j(2gt) t^{j+1} dt, \quad (16)$$

$t_1 = R_1/a_B$, $t_2 = R_2/a_B$, $K_j(x)$ is the $j$-th order ($j = 0,1$) modified Bessel functions of the third kind,

$$\mathbf{n} = \frac{1}{2}\left(\frac{a_B}{a_c}\right)^2 \equiv \frac{H}{2H_0}. \qquad (17)$$

Here the magnetic field characterizing value $H_0$ is determined by the parameters of the wire substance:

$$H_0 = \frac{m_1^2 e^3 c}{c^2 \hbar^3}. \qquad (18)$$

For the *GaAs* wire $H_0 \approx 6$ T.

## 4. Discussion of results

In numerical calculations carried out for the *GaAs* wire coated by the $Ga_{1-x}Al_xAs$ layer, following values of parameters have been used [11]: $E_R \approx 5.2$ meV, $a_B = 104$ Å, $m_1 = 0.067 m_0$, $m_2 = (0.067 + 0.083x)m_0$ ($m_0$ is the free electron mass) and $V_0 = 1.247 x Q_e$ eV ($Q_e$ is the conduction-band discontinuity fraction) for variation of the alloy concentration $x$ within the limits $0 \leq x \leq 0.45$.

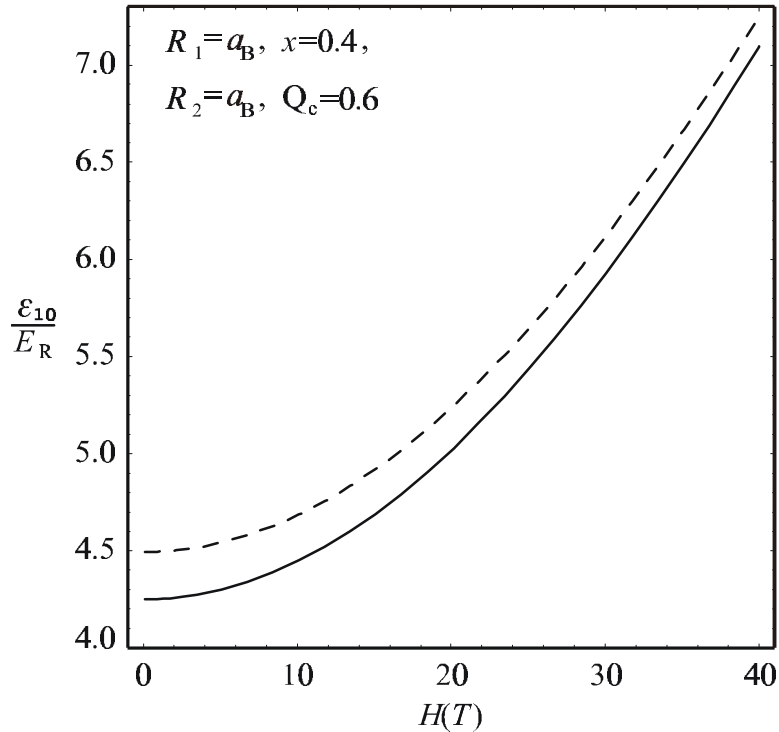

Fig. 1. The ground energy level $\mathbf{e}_{10} = E_{100}$ dependence on magnetic field is presented (doted line- $m_1 = m_2$, solid line- $m_1 \neq m_2$).

It should be note that the applicability of the effective mass method in the $GaAs$-$Ga_{1-x}Al_xAs$ system is connected with the clarifying of the role of the Ã-Õ mixing. According to [12] for V- and T-shape 1D-systems the mentioned effect begins to play a decisive role for the values $R < 50 \text{Å}$ and $x > 0.5$.

In Fig. 1 the dependence of the ground energy level $e_{10} = E_{100}$ on magnetic field is presented. As is seen from Figure, $e_{10}$ does not depend on magnetic field up to the value $H_0$, i.e. for $H \leq H_0$ the size–quantization plays the main role ($R_1 < a_c$). As it follows from calculations, in the $m_1 = m_2$ approach (dotted line in Fig. 1), the electron energy is greater than when the mass difference is taken into account.

As was expected, with increasing of magnetic field, i.e. with decreasing of the electron localization region, the difference $e_{10}(m_2 = m_1) - e_{10}(m_2 \neq m_1)$ tends to zero, what, in its turn, means that electron does not "feel" the presence of the barrier layer. According to calculations, the states with $n = 2$, $n = 3$ are more sensitive with respect to the increasing of $H$.

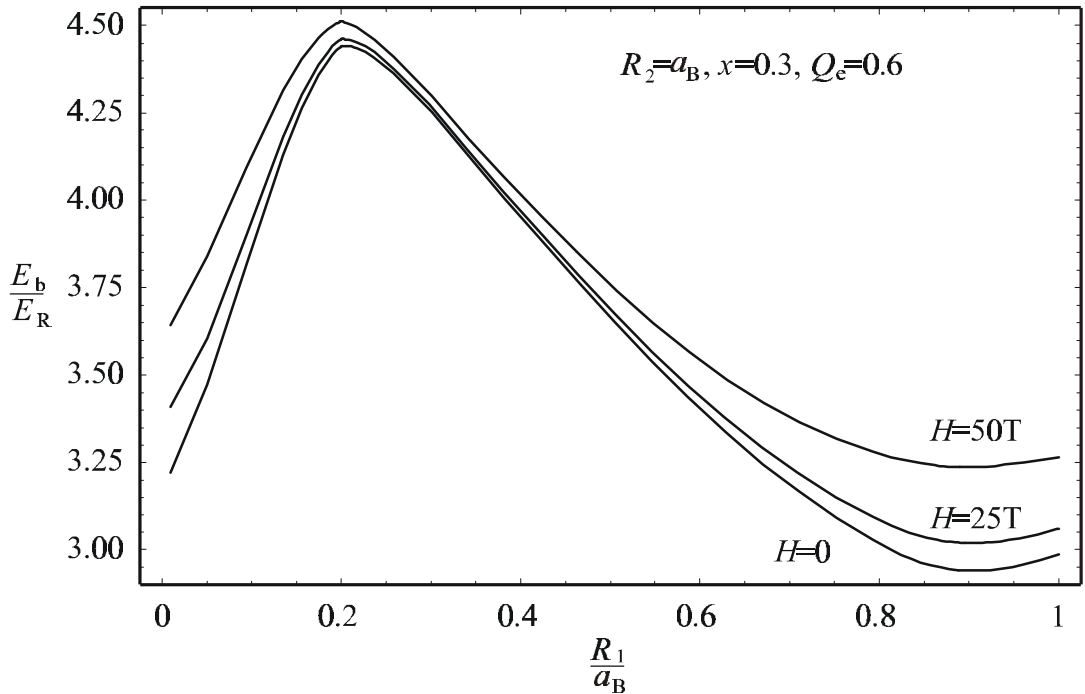

Fig. 2. The binding energy dependence on the wire radius for various values of magnetic field.

Also the dependencies of levels $e_{n0}$ (for $n=2$ and $n=3$) on the wire and coating radii for given $x$, $Q_e$ and $H$ have been studied. It is found that the $e_{n0}(R_1)$ and $e_{n0}(R_2)$ dependencies at $H \neq 0$ have the same peculiarities as for $H=0$ [8] (particularly the peculiarity connected with mass difference in the wire and coating).

In Fig. 2 the binding energy dependence on the wire radius for various values of magnetic field is presented for $x=0.3$, $R_2=a_B$, $Q_e=0.6$, $m_2 \neq m_1$ (without regard to Ã-Õ mixing). With the increase of $H$ the binding energy increases quickly within the region $R_1 \leq 0.1a_B$ and $R_1 \geq 0.6a_B$ and slowly in the region $0.2a_B \leq R_1 \leq 0.5a_B$. Such a behavior of the binding energy is the consequence of the fact that for $0.2a_B \leq R_1 \leq 0.5a_B$ the size-quantization prevails the magnetic one. The minima of the binding energy for various $H$ at $R_1 \approx 0.9a_B$ are connected with the electron mass difference in the wire and coating [8]. With increase of $H$ the depth of minimum becomes smaller and tends to zero because the strong field localizes electron in the near axis region.

For the fixed value of $R_1$ the $E_b$ value passes maximum at $R_2 = R_1$ and then falls abruptly, tending to the value obtained in Ref. [7] (i.e. to the limiting value for $R_2 \to \infty$). The curves corresponding to the large values of $H$ decrease relatively quickly.

In Fig. 3 the impurity center binding energy dependence on magnetic field is presented for various parameters values of the problem. From comparison of curves 1, 4 and 6 it follows that (for given $R_2 = 1.5a_B$, $x=0.3$, $Q_e=0.6$) the binding energy increases with the decrease of the wire radius, and thereby the rise velocity depending on $H$ is higher for larger values of radii. From comparison curves 2, 3 and 4 it is seen that (for given $R_1 = 0.75a_B$, $R_2 = 1.5a_B$, $Q_e=0.6$) the binding energy rises with the increase of the concentration $x$. Thereby the rise velocity depending on $H$ is larger for small values of the concentration. The mass difference role in the wire and coating one can see from the curves 4 and 5. From calculations follows that, even for $H=40\,\text{T}$, the error entering by $m_2 \approx m_1$ approach is the same as for $H=0$.

From model calculations it follows that the curves 2, 3 and 4 converge at $H=80\,\text{T}$, and all curves converge at $H=150\,\text{T}$ what means that magnetic quantization prevails over the size quantization.

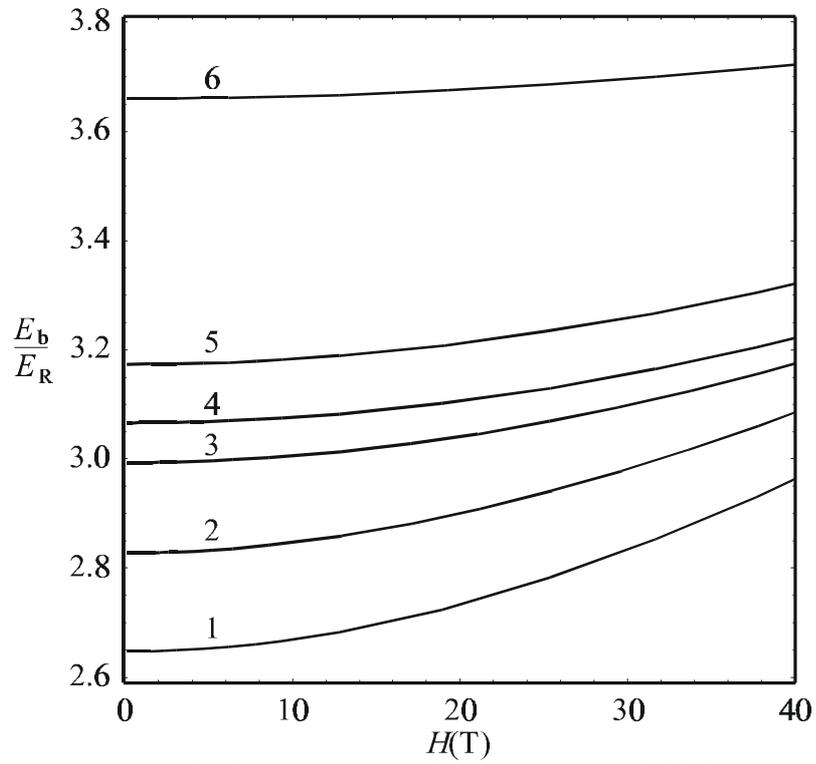

Fig. 3. The impurity center binding energy dependence on magnetic field for various parameters values of the problem: $R_2 = 1.5a_B$, $Q_e = 0.6$; 1. $R_1 = a_B$, $x = 0.3$. 2. $R_1 = 0.75a_B$, $x = 0.1$. 3. $R_1 = 0.75a_B$, $x = 0.2$. 4. $R_1 = 0.75a_B$, $x = 0.3$. 5. $R_1 = 0.75a_B$, $x = 0.3$, $m_2 = m_1$. 6: $R_1 = 0.5a_B$, $x = 0.3$.

8 Janurary 1999                                                                                           Yerevan State University